\documentclass[11pt]{article}
\usepackage{geometry}                
\usepackage{graphicx}
\usepackage{amssymb}
\usepackage{caption}
\usepackage{subcaption}
\usepackage{amsmath}
\usepackage{siunitx}
\usepackage{fixmath}
\usepackage{MnSymbol}
\usepackage{amsfonts}
\usepackage{array}
\usepackage{multirow}
\DeclareMathAlphabet{\mathpzc}{OT1}{pzc}{m}{it}

\usepackage{caption}
\usepackage{subcaption}

\title{Inconsistencies in the Notions of Acoustic Stress and Streaming}
\author{Clifford Chafin\\\ \small{Department of Physics, North Carolina State University, Raleigh, NC 27695} \thanks{cechafin@ncsu.edu}}

\begin{document}
\maketitle

\begin{abstract}
Inviscid hydrodynamics mediates forces through pressure and other, typically irrotational, external forces.  Acoustically induced forces must be consistent with arising from such a pressure field.  The use of ``acoustic stress'' is shown to have inconsistencies with such an analysis and generally arise from mathematical expediency but poor overall conceptualization of such systems.  This contention is further supported by the poor agreement of experiment in many such approaches.  The notion of momentum as being an intrinsic property of sound waves is similarly found to be paradoxical.  Through an analysis that includes viscosity and attenuation, we conclude that all acoustic streaming must arise from vorticity introduced by viscous forces at the driver or other solid boundaries and that calculations with acoustic stress should be replaced with ones using a nonlinear correction to the overall pressure field.  
\end{abstract}

This article is meant as the fourth part of an installment on wave motion \cite{Chafin-EM, Chafin-rogue,Chafin-cd}.  One of the major thrusts of 20th century physics has been unification.  Through some combination of vanity and poor conceptualization in favor of clever mathematical presentation we have often over reached in this respect.  Aspects of different wave systems are attempted to be treated by similar frameworks when, given more careful analysis, no such unified description is justified.  One of the greater sins has been in the use of pseudomomentum \cite{McIntyre}.  Translational symmetries in systems give conserved quantities that have the units of linear momentum yet have no such meaning in terms of actual forces at boundaries.  Often these arise from clever integration by parts constructions where the integration constant is ignored.    In the case of fluids, these translational cases often miss essential nonlinear features of packet motion.  When using conserved quantities, especially angular momentum, the end-of-packet contributions can be dominant.  Since conservation laws give a powerful check on often freewheeling perturbation approaches, getting this right is very important.  

In the case of dielectric response, the Abraham-Minkowskii debate has lingered over a century.  It can be considered as either about the correct momentum of a photon in a medium or as the correct electromagnetic stress tensor for the medium containing a photon.  These then are used to determine the forces at the surfaces of media under reflection, transmission, etc.  Averaging methods get very complicated and leave room for doubt as to the validity of approximations.  It is argued that proper use of boundary conditions make both theories equivalent \cite{Pfeifer}.  By constructing a simpler dielectric model that lets the radiative field and medium response always be decomposable \cite{Chafin-EM}, we can give exact answers for how the energy and momentum are shared and what the relevant surface forces are.  Interestingly, this gives some unexpected end-of-packet contributions as well and all by elementary means.  Rather than asking what the detailed decomposition is in any particular dielectric, this gives a set of universal results for these forces independent of their particular construction.  Constitutive relations are seen to be due to a particular subset of branches of the totality that give any possible field and media initial data.  

Stress in this system is caused by the radiative stress of the electromagnetic fields that are absorbed and emitted by the medium.  The two component nature of this system, charged solid components with elastic forces and transiently free electromagnetic waves allow the medium to experience a true stress rather than a pressure.  In hydrodynamics, this is not the case as there is only a massive liquid and, in the inviscid limit, a pressure to drive behavior.  This is true whether or not there is an oscillatory component to the pressure.  

In the case of ocean waves, there is a large literature on ``wave stress'' and ``wave action.''  The first notion is somewhat distressing in that inviscid simple fluids are stress-free by definition.  Reynolds' stresses for periodic motions are in the category of pseudo-forces that need not give real impulses on boundaries and inclusions.  These, no doubt, have inspired an uncritical faith in such calculations.  Hydrodynamics is distinct from most other field theories in that it contains and intrinsic nonlinearity.  This is not just an interaction term with other fields that can be viewed as nonlinear in some perturbative scheme.  It is a fundamental nonlinearity associated with advection.  Stokes expansions give a way to modify periodic solutions to give new ones in the nonlinear regime but it does not help with packet motion.  Generally we assume that knowing the infinite wave solutions we can derive anything about packets from them.  In this case, this assumption is false \cite{Chafin-rogue}.  The nonlinearities can create packet length surface elevation changes that allow the Stokes drift to stay with the support of the packet in an incompressible fluid (where elevation changes must accompany any net transport in mass).  Wave action turns out to be equivalent to the angular momentum density.  End-of-packet variations can dominate most angular momentum effects so this turns out to be of questionable value as a locally conserved quantity.\footnote{Harmonic solutions with translational symmetry often end up with hidden contributions to conserved quantities at infinity \cite{Chafin-EM}}  

Wave-stress \cite{LH} is used to predict wave set-up on beaches and forces on submerged and floating objects.  This has never been an overwhelming success.  Given the irregularity of real waves this is not surprising but is is certainly disappointing.  Wave behavior in wave tanks and the ocean can differ markedly.  One can dispense with wave stress in terms of a time average ``pressure charge'' term $\rho_{P}=-<\rho \partial_{ij}\phi\partial_{ij}\phi>-<\rho\partial_{t}\nabla^{2}\phi>$ where $\phi$ is the velocity potential.  The long range pressure field is then given by 
\begin{align}\label{P}
\nabla^{2}P=\rho_{P}=-<\rho \partial_{ij}\phi\partial_{ij}\phi>-<\rho\partial_{t}\nabla^{2}\phi>
\end{align}
where $\rho_{P}$ is typically a purely positive definite source. 

The second term typically makes no contribution to lowest nonzero order.  Pressure in incompressible fluids play the role of a Lagrangian constraint.  If it depends on some time dependence in a circular manner then we are faced with an iterative problem.  The free surface does introduce some complications analogous to the image charge problem in conductors \cite{Chafin-rogue}.  Implementation of this procedure gives long range and surface elevation effects at packet reflections not found in the wave stress approach.  Additionally, the transverse forces cannot be given by distinct stresses $S_{yy}\ne S_{xx}$ as argued by Longuet-Higgins \cite{LH} but by a global pressure field that varies continuously at the edges of the container.  

Acoustic waves have some similarity to surface waves.  They are both governed by the equations of hydrodynamics.  Acoustic waves require some compressibility but usually this is a small quasiperiodic variation of bulk flow.  Surface waves require an external gravity-like field to produce an equilibrium at a, usually sharply defined, surface.  Surface waves transport mass by irrotational motion through this free surface deformation.  Acoustic waves typically arise with unbounded boundaries or in a container with rigid ones.  These can be in the form of acoustic beams that fill the container or a narrow channel of it.  The ``container'' can be a liquid held by gravity in an open vessel where the impedance mismatch at the surface keeps the waves confined to it by total reflection.  

Two common questions arise in such situations: ``How does sound generate forces on the container?'' and ``What is the fluid response to attenuation of the wave?''  The first case is the topic of acoustic stress and the second of acoustic streaming.  For similar reasons as above we have reason to be skeptical of stress based descriptions of forces on boundaries.  In the case of attenuation, damping is often due to some viscosity which is an avenue for vorticity to enter the fluid which is necessary for a flow to be initiated in a filled container with fixed boundaries.  We also will have reason to reconsider some of the reasoning underlying common explanations of acoustic streaming.  

First we will consider the case of standing sound waves in containers and the forces they exert on boundaries.  Through a set of consistency arguments we will demonstrate (redundantly) that acoustic stress is a flawed concept.  Following this we will consider the case of acoustic streaming including the notion that sound carries momentum.  While this is true as a tiny relativistic correction, $p=\frac{E}{c^{2}}v_{g}$, we demonstrate that this is paradoxical in any measurable/nonrelativistic sense.  We compare this with surface waves where such a momentum density is nonzero.  We discuss the ways boundary effects can lead to vorticity sources that convert wave motion to net streaming flow and how, absent such boundaries, attenuation of sound can create only irrotational forces, hence, no streaming flows.  

\section{Acoustic Forces}
The forces that sound can exert will be shown to be either boundary forces or long range elastic corrections that, for liquids, tend to be small density changes.  Both are mediated by pressure in the inviscid limit.  In gases, sound waves require thermodynamic corrections.  Isaac Newton was not aware of this and this resulted in his predictions for sound speeds in air to be off by 15\% in some cases \cite{Newton}.  When thermodynamics was developed this was rapidly corrected.  In the case of liquids, the compressibility is not very sensitive to temperature as almost all the energy is from compression of bonds and orbitals.  This makes the analysis in this case simpler so we restrict our discussion to the case of liquids.  

\begin{figure}[!h]
   \centering
   \includegraphics[width=4in,trim=0mm 60mm 0mm 120mm,clip]{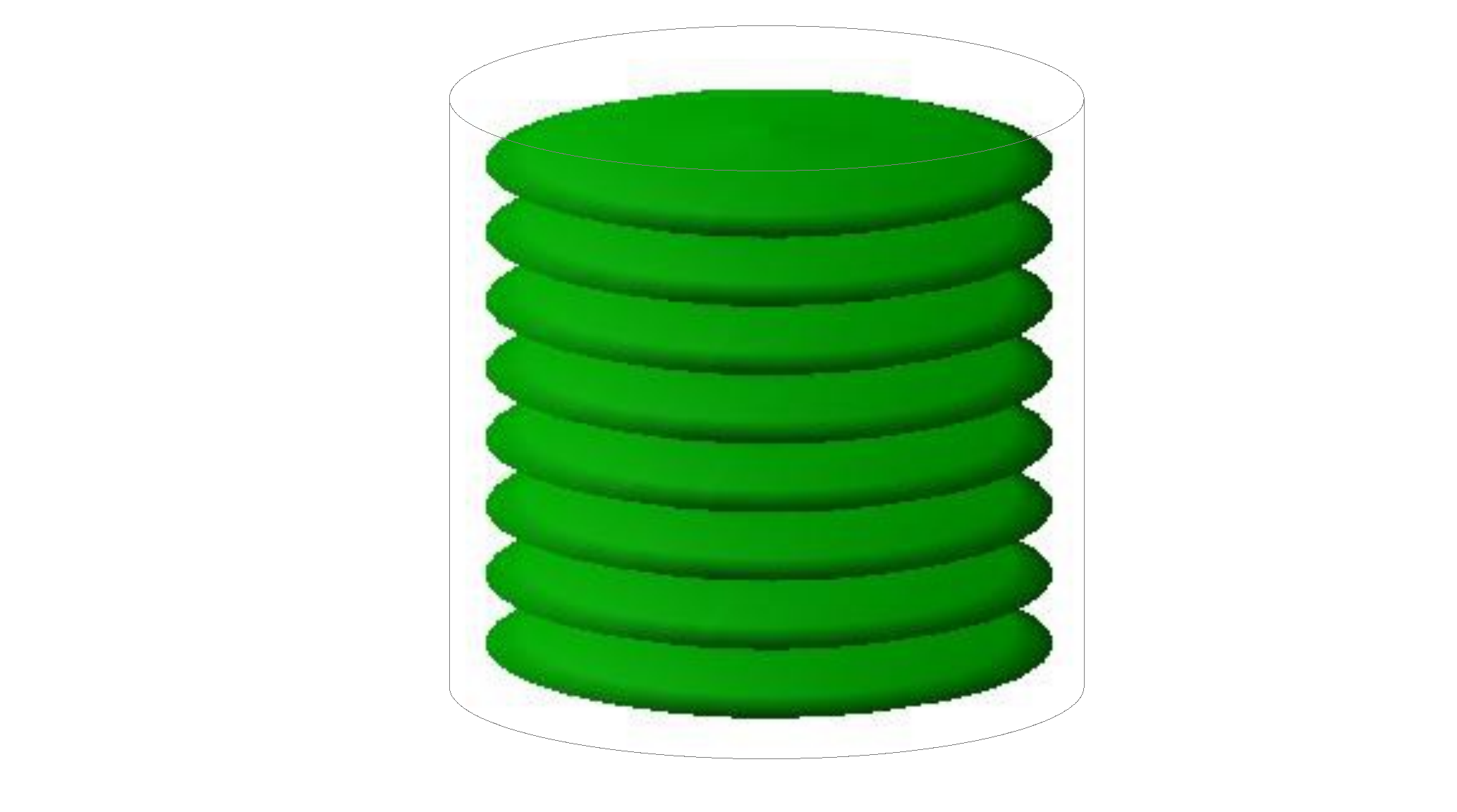} 
   \caption{Standing sound wave in a closed cylindrical tube.}
   \label{tube}
\end{figure}

Consider the case of a standing wave in a tube as in fig.\!~\ref{tube}.  Assume that the tube was initially filled with a liquid at constant pressure and the walls are rigid then sound waves are generated by actuators than span the ends.  
No slip effects at the sides will give some damping in a narrow layer at the sides.  Since the displacements of acoustic fields are usually very small, this layer $\delta$ will typically be much less than the wavelength $\lambda$.  The local particle motion in the acoustic wave is given by $A\cos(kx-\omega t)$ so that the long range pressure field is a solution to $\nabla^{2}P=c=\frac{1}{2}\rho A^{2} \omega^{2}k^{2}\sim \mathcal{E}/\lambda^{2}$ when averaged over a wavelength.\footnote{One could consider the waves to have a lowest mode radial component so $\nabla^{2}P=c\sin(r/R)$.  Here we consider the radial component to be more flat with steep attenuation at the walls for simplicity.}  In the case of a very long tube with radius $R$, the instantaneous pressure can be solved as 
\[
P=P_{0}+\frac{c}{2}r^{2}
\]
The pressure imbalance will soon be equilibrated by radial density changes.  In a liquid such changes are so small as to be essentially isothermal so that the compressibility can be expressed as $\beta=\frac{1}{\rho v_{s}^{2}}$.  The correction to the density can be expressed as $\delta\rho=\beta\rho_{0}\delta P$. 
 The equilibrating density becomes $\rho(r)=\rho'-\frac{1}{4}\rho_{0} A^{2}k^{4}r^{2}$ where  $\rho'=\rho_{0}(1+2A^{2}k^{4}R^{2})$ is fixed by mass conservation.  For a finite tube there are corrections to this for the field within one radius of the tube.  

If we drive a beam of the same radius (neglecting spreading) through an infinite body of water the solution is
\[
P=
\begin{cases}
   P_{0}+\frac{c}{2}r^{2} & \mbox{when } r<R\\
    P_{0}-\frac{c}{2}R^{2}(\ln(r/R)-1) & \mbox{when } r>R
\end{cases}
\]
where we have fixed the pressure to the the same at the center.   
The low compressibility of the fluid ensures that the density changes and net displacements to do this will be small.  In fact, an irrotational force on a fluid in a container with fixed boundaries can induce only oscillations but no net circulation so that damping leads to equilibration of the fluid outside the support of continuously driven waves.  
Note that the forces on the container are given entirely by a pressure not a stress.  The short time scale oscillations exist in the support of the waves but these longer range pressures become more static as we get farther from this region analogously to the long range pressure distortions from wave packets of surface waves \cite{Chafin-rogue}.   

\begin{figure}
        \centering
        \begin{subfigure}[b]{0.3\textwidth}
                \includegraphics[width=\textwidth]{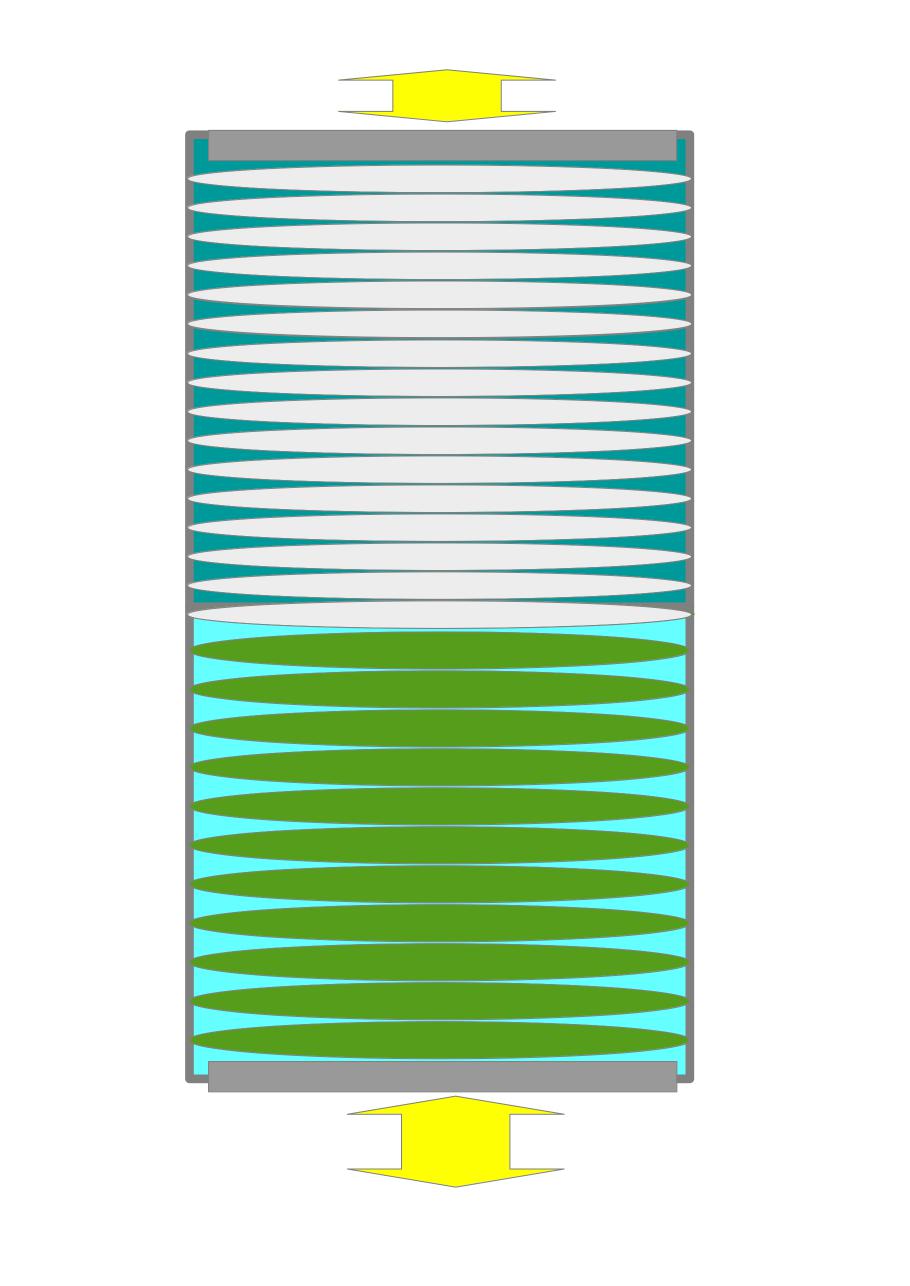}
                \caption{Standing sound wave in a cylindrical tube of two immiscible liquids.}
                \label{wide}
        \end{subfigure}%
        ~ 
        \begin{subfigure}[b]{0.3\textwidth}
                \includegraphics[width=\textwidth]{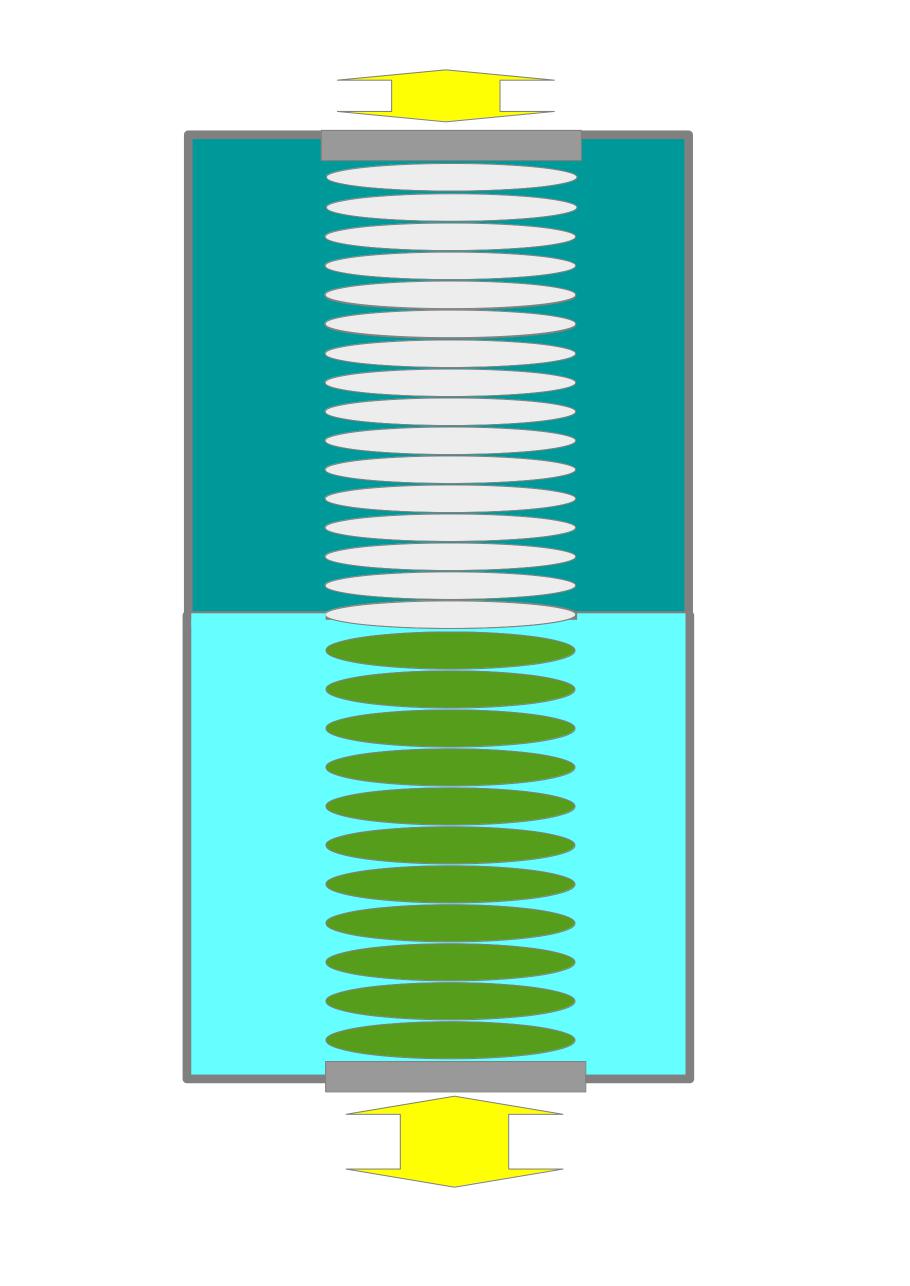}
                \caption{Narrow beam of sound in the same tube with an immiscible interface.}
                \label{narrow}
        \end{subfigure}
        \begin{subfigure}[b]{0.3\textwidth}
                \includegraphics[width=\textwidth]{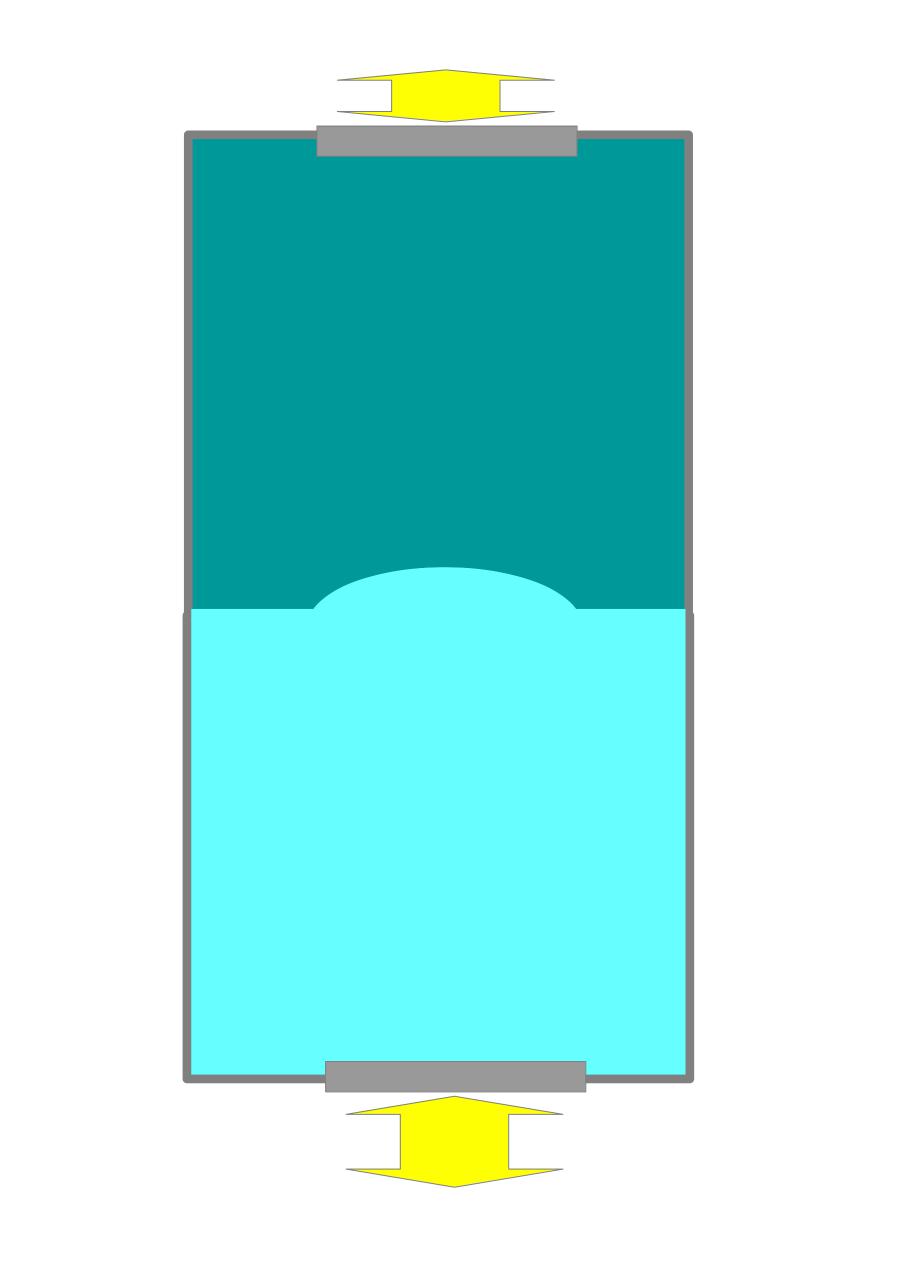}
                \caption{Surface elevation changes from sound waves at interface.  }
                \label{bulge}
        \end{subfigure}
        \caption{}
\end{figure}

Now consider the case of a tube with two different fluids as with immiscible fluids on top of each other.\footnote{For a narrow tube example, see app.}  This is the type of configuration that is used to measure acoustic stress by observing the deformation of the free surface \cite{Beyer}.  The case of a beam that spans the tube is given in fig.\!~\ref{wide}.  A narrow beam case is given in fig.\!~\ref{narrow}.  
We have deliberately included drivers at both ends of the tube.  If we send waves from one side then there is some reflection at the interface and the transmitted wave then reflects at the other.  By having a second actuator we can control the relative phase at both ends and optimize the energy density of standing wave components in each fluid region and avoid any reflection or wave damping at the ends.  

It is known that such configurations lead to a bulge or depression at the free surface \cite{Beyer} as indicated in fig.\!~\ref{bulge}.  There are two ways this can arise.  First, vorticity production at the driver can induce a net circulating flow in the narrow beam case.  The impulse when such a flow reaches the surface will lead to deformation.  In the case of a beam only driven by the lower driver, this will always be upwards.  Since we know that this is not always the case in experiments, this cannot be the only effect.  

The energy density in the upper and lower regions need not be the same.  Amplitude and frequency of motion must match at the interface but the size of the oscillations and the density can vary in each.  Any net elastic restoring of pressure fields is assumed to have already occurred.  However, this system does not have translational symmetry.  The density corrections in each half will generally be different and there will be some horizontal interfacial forces to account for.  

{\bf Case 1:} First consider the case of fig.\!~\ref{wide} where the beam spans the tube.  Asymptotically the net pressure averaged across a cross section is the same for both fluids.  The one with the larger energy curvature, $c$, value will have pressure depletion at the center of the beam and pressure excess at the edges relative to the pressures of the lower $c$ fluid region.  This indicates the fluid will bow toward the larger energy curvature region.  

{\bf Case 2:} In the case of a narrow beam relative to the container radius in fig.\!~\ref{narrow}, we have a long logarithmic density tail correction.  This ensures the higher energy curvature fluid region will have a net higher pressure correction in the region around the beam and so lead to a bowing of the surface in the opposite direction.  Quantifying these effects is difficult as they are strongly dependent on the geometry of the beam and container.  The logarithmic nature of the radial field means that the container boundaries can play an important role unless they are many times farther away than the beam radius.  If we consider a long thin beam then the correction to the pressure looks more like a charged plate than a wire and the corrections have no bound on the range of their effect.\footnote{Interestingly, some surface deformations are much more intricate and irregular that a simple stress picture would suggest \cite{Beyer}.  The more structurally interesting nonlinear pressure field presented here seems a promising candidate to reproduce them.}  

These pressure fields that extend beyond the support of the beam and drive the surface deformation can be thought of as ``nonconstitutive'' in the sense that the medium response is not naively a function of the local motions that a stress based approach assumes.  Similar problems arise in electromagnetism where forces on wires are not a simple function of the current there but of long range electric fields created by polarization of the wire itself \cite{Rosser}.  Of course, all physics is local but sometimes we neglect long range effects in local analyses of the most evident quantities.  

Experimental results for acoustic stress have never been very consistent with theory and sometimes even each other.  This analysis my point the way to some resolution.  Some may argue that the perturbative approach is almost certainly right based on the duration of its acceptance and the stature of those who derived it.  In response I argue that if current perturbative models and interpretation of acoustic stress are correct they should agree with purely pressure based analysis and other theorems such as vorticity transport and energy and mass flux conservation at the driver.  This alone justifies such an alternative treatment.  There is certainly room for a more precise pressure based analysis of the acoustic response than has been done here but agreement at this point seems very unlikely.

\section{Acoustic Streaming}
\subsection{Paradoxes}

There are multiple kinds of acoustic streaming but all have in common that sound waves drive motion of the ambient fluid with it \cite{Gedeon, Lighthill}.  This can be considered a number of ways.  Some of them will be shown to be inconsistent.  Let us begin with a comparison with surface waves. The linearized purely irrotational solutions to surface waves allow a net forwards drift by virtue of the time changing surface.  In the case of finite depth, the solutions give vanishing advance of particle motion at the bottom.  This defines a preferred reference frame for the fluid.  We can superimpose flows and waves to some extent.\footnote{See Chafin \cite{Chafin-rogue} for restrictions on this.}  If we neglect drag on the bottom, the whole fluid mass can translate and carry surface waves on it.  This alters the dispersion relation since the frequency of waves of the same wavelength is different.  Superimposing such bulk translational flows without waves is harmless as we just get a new net translation of the body of liquid \cite{Lighthill78}.  However, if we try to superimpose waves on different flows there is a problem.  The resulting solution does not evolve as a superposition of the independent ones.  

In the case of sound waves in a long tube as in fig.\!~\ref{tube}, there is no reference frame but the wall.  Ignoring friction or viscous effects there that may take a long time to propagate into the center of the beam, there is no way for the system to determine a special frame except for the mean motion of the particles.  For aperiodic solutions, this may have no meaning at all.  There are attempts to discuss acoustic streaming by finding nonlinear corrections to the linear wave solutions that give a net translational motion \cite{LL, Gedeon}.  Perturbative means can give such results but we must then ask if the results are meaningful.  
There are several arguments that show that the notion of sound carrying momentum is inconsistent.  Consider the following thought experiments.  

{\bf Example 1:}  A finite tube with oscillators at each end are phase matched so that a propagating wave moves from one to the other with no reflected wave.  We are then inclined to call one the ``emitter'' and the other the ``absorber.''  Consistently, these give sources and sinks of the wave energy.  This can be thought of as the case in fig.\!~\ref{wide} with a single fluid component and a different driving pattern of the actuators.  The energy imparted to the wave can be directly evaluated by $F\cdot v$ at the oscillators.  Certainly no mass is transported.  This could be altered by using a porous actuator that injects fluid with its motion.  Otherwise, if the sound waves carry mass then there must be a pressure gradient across the beam to halt it.  This either distorts the solutions to eliminate mass flux or forces a destruction of the wave to create an ``Eulerian'' backwards flow.  We reject the later notion since it would introduce an extra energy sink that we have chosen the driver motion to prevent.  

{\bf Example 2:}  Another example is currently employed with cold gas traps.  Optical beams can create a annular channel which hold certain hyperfine states of ultracold atoms \cite{Dolfovo}.  These can be highly quantum degenerate but this is not necessary.  Detuned laser beams can excite sound waves that traverse such a circuit with no physical walls to exert damping or to define a particular reference frame.\footnote{One can always argue that rotation defines a reference frame but in the large curvature limit this argument is sufficient for our purposes.}  If sound waves carry mass then what if we simply give counter rotation so that the net mass flux is zero?  How are we to say that this is not the correct frame for such a solution to begin with?  If a solution has a net mass flux, this momentum must come from the driver.  Since there are solutions with sound with no net flux, then there are presumably local driving forces that drive a packet without disturbing the bulk motion of the fluid.  In the case of surface waves, deep packet scale pressure fields uplift the packet so that mass carried by the Stokes drift can be carried with the support of the packet.  In the case of an acoustic tube there is no deep fixed reference frame.  The only way a packet could carry net mass would be for it to become more dense than its surroundings.  To mediate such propagation there need to be end-of-packet inwards forces and some of the particles must propagate at the group velocity or faster.  For sound waves in a highly correlated liquid this is impossible.  Deep water surface waves do not have this restriction because the net flow can be made up by deep motions at very slow velocity $\sim a\omega\ll v_{g}=\frac{1}{2}\sqrt{g/k}$.  

\begin{figure}[!h]
   \centering
   \includegraphics[width=3in,trim=20mm 180mm 30mm 90mm,clip]{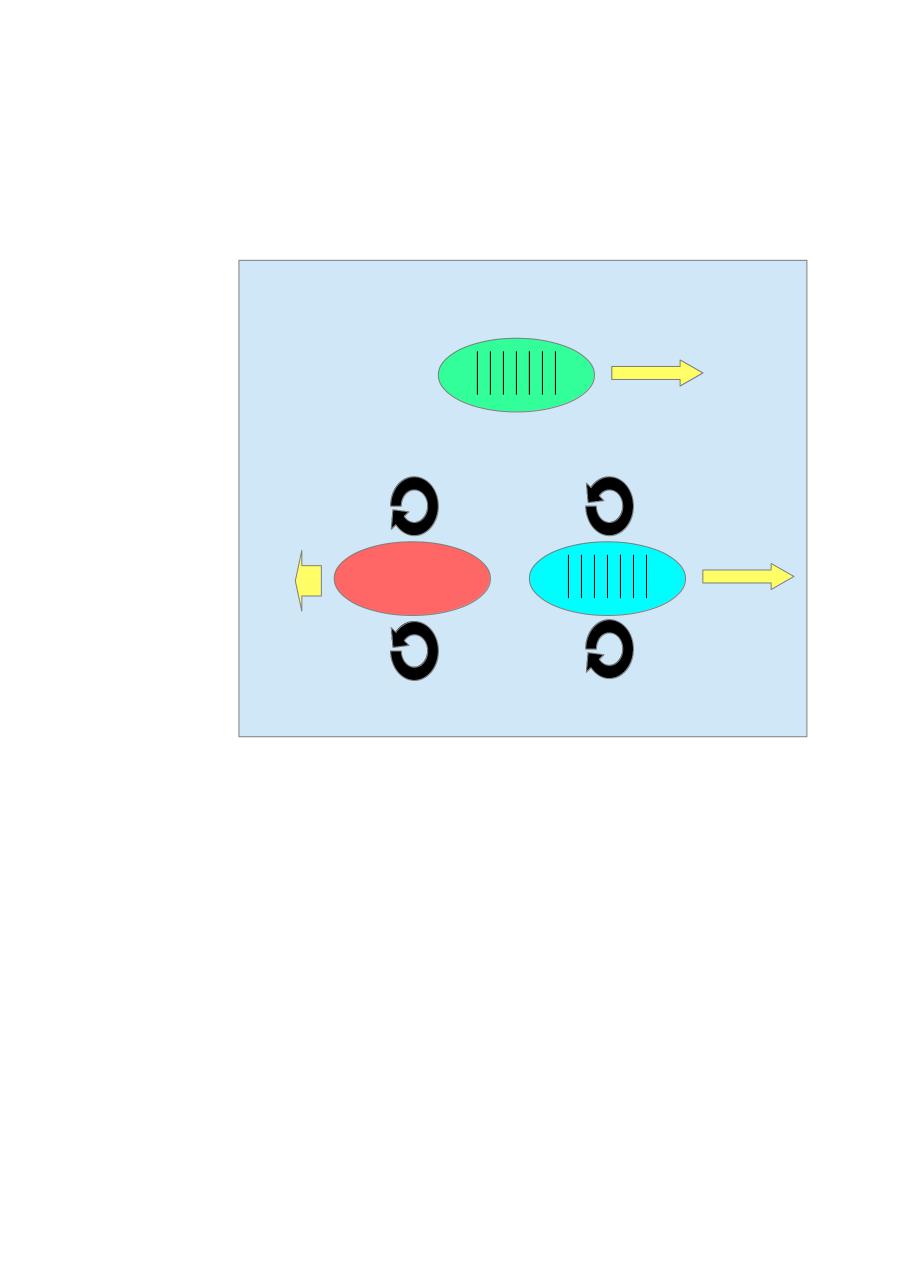} 
   \caption{A packet with local canceling Eulerian flow undergoing separation and the creation of vorticity.  }
   \label{vorticity}
\end{figure}

{\bf Example 3:}  Consider the case of a localized sound packet of length $L$ in an infinite medium as in fig.\!~\ref{vorticity}.  If such a solution carries mass and we supply a local counterflow to cancel it then when the packet moves off such a region of (relatively slow) counterflow support it leaves behind a backwards moving region.  The initial data has no vorticity when averaged over the time scale of period oscillations.  As the packet and counterflow separate they both acquire vorticity at the boundaries of packet support.  Vorticity conservation requires that vorticity be created only by viscous forces or vortex stretching.  Since neither of these occur here, the notion of a sound beam carrying intrinsic momentum is inconsistent.  

Let us now review some of the reasons people suspect that sound waves might carry momentum as an intrinsic property.  Perturbative corrections \cite{LL} give a nonzero momentum flux density.  These sort of arguments tend to gloss over the effect of forming packets and assume they have negligible edge effects.  Two problems exist here.  Firstly, building packets as superpositions of waves that have already had nonlinear corrections surely has further corrections.  Secondly, as we have seen in the case of ocean waves \cite{Chafin-rogue} hydrodynamics can give long range nonlinear corrections to the pressure that makes such an analysis have some inconsistencies.  Computations of wave stress typically involve some integration by parts of segments of infinite waves \cite{Lighthill78} with an integration by parts and association of the integrand directly with true (not pseudo) momentum.  Careful analysis in terms of fixed boundaries or packets show that this is not consistent with a fully pressure based description of forces at boundaries \cite{Chafin-rogue}.  

Calculations from QFT involve quasiparticles interacting in diagrams that conserve momentum.  In reality these are purely pseudomomentum conservation laws.  In the case of electrons in a metal, the group velocity is the relevant velocity for the transport of mass, as made clear in semi-classical electron calculations.  When one considers the normal modes of a solid, these get quantized but are strictly the relative modes and do not include the net translational motion.  Sometime this is artificially attempted to be included as a ``zero mode.''  Only the translational mode actually carries momentum.  The rest are defined in the material of the solid in which it is at rest.  Translation alters the dispersion relation by altering the frequency as a function of the boost.  This point is sometimes obscured because of experiments with superfluid Helium where are low energy oscillations with a phonon-like dispersion relation that do carry mass.  These are the reason sound and heat flux in superfluid Helium can carry mass and how the ``viscous'' transfer of momentum in the Andronikashvilli experiment between plates can exist when no normal component exist.  

In hindsight, we should really view the translational states from perturbation theory that contain momentum as including a background boost in the underlying fluid.  As such, these are not really superimposable as discussed above.  This is not simply a matter of a problem with superimposing waves that have had a nonlinear correction but of one of superimposing waves that have a different underlying characteristics (in this case translational velocity).

\subsection{Attenuation and Vorticity}
Some treatments of acoustic streaming \cite{Lighthill} employ acoustic dissipation as a mechanism for driving the flow.  Consider a vessel of fixed size and shape containing a liquid of low compressibility.  If the fluid contains no vorticity then the velocity field is given by $v=\nabla\phi$.  In a very short time a pressure field arises to cancel any slow net velocity field so that bulk time average motion is impossible.  To achieve bulk motion, vorticity must be created.  This is certainly possible at the driver.  There is an obvious ``pushing-sucking'' asymmetry at the edge of the driver that allows fluid to flow in during the back stroke and get advanced during the forwards stroke.  Damping of forwards and backwards oscillation of fluid motion can give a net flux if it is damped in a fashion that is directionally biased.  
It is the author's opinion that all acoustic streaming is generated in this way.

\begin{figure}[!h]
   \centering
   \includegraphics[width=3in,trim=20mm 240mm 30mm 10mm,clip]{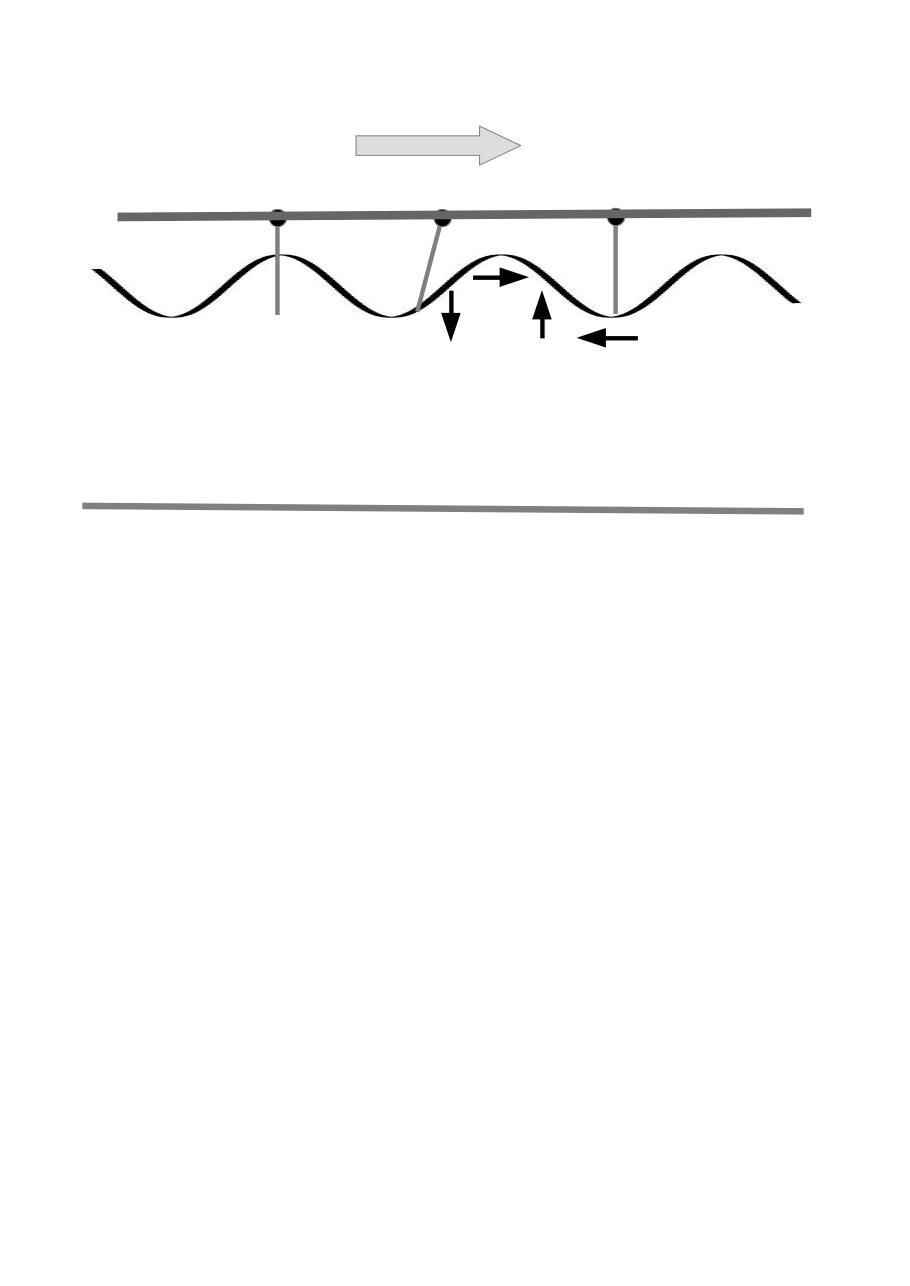} 
   \caption{Half swinging wave dampers.  }
   \label{ratchet}
\end{figure}

As an example of how aperiodic boundary effects can lead to a net flow consider a surface wave.  The momentum of surface waves is relatively small, $\frac{\mathcal{E}}{p}=\frac{\omega}{k}$.  The mass drift is relatively small considering the considerable mass involved in the wave motion.  The energy in deep water waves is evenly divided among the kinetic and potential energy with the kinetic motion from nearly circular particle orbits.  
Consider a wave damping ratchet mechanism for converting the kinetic motion into flows.  A rack of pivoting plates sits above the waves as in fig.\!~\ref{ratchet}.  The wave is propagating the the right or left.  As the fluid flow advances to the right the ratchet locks in place and converts most of the kinetic motion that impinges on it to turbulence with little net resulting momentum.  At the surface the kinetic motion has velocity $a\omega$.  One pass of the wave removes this kinetic energy per area $\sim \frac{1}{4}\rho (a\omega)^{2}\cdot 2a$ in time $T=\frac{2\pi}{\omega}$.  The remaining energy is in a standing wave where there is no net motion and the motion at the troughs are now purely vertical and there is no net momentum.  We see that the fractional change in energy of the wave is 
\begin{align}
\frac{\frac{1}{2}\rho (a\omega)^{2}a}{\frac{1}{2}\rho g a^{2}}=\frac{ a\omega^{2}}{ g }={a k}
\end{align}
The momentum of the Stokes drift has now been converted transferred to the plates and to a recoil flow in the opposite direction to the waves.  

\begin{figure}[!h]
   \centering
   \includegraphics[width=3in,trim=10mm 140mm 30mm 170mm,clip]{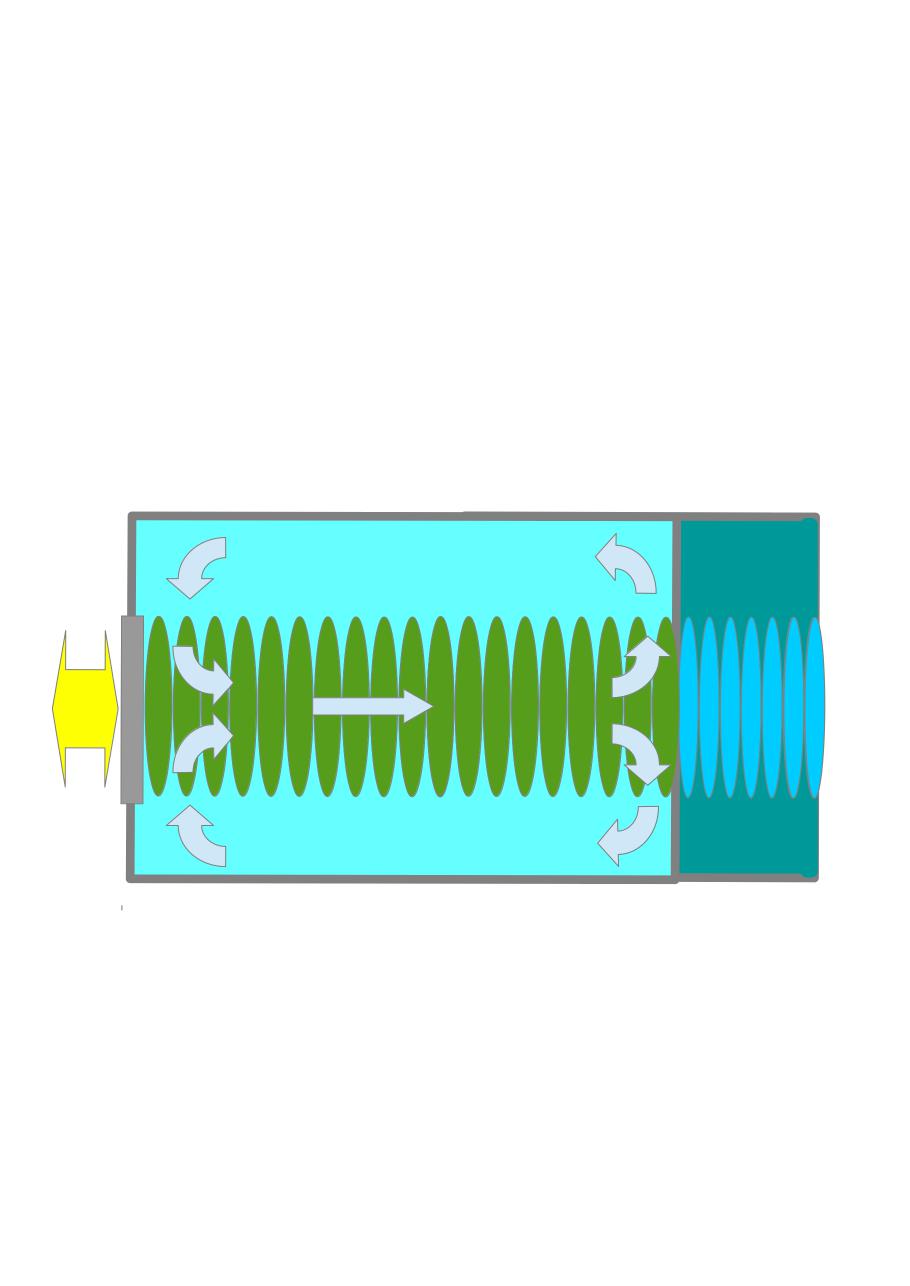} 
   \caption{Sound and mass flux.}
   \label{flux}
\end{figure}

In examples of acoustic streaming one sees a growing jet forming at the driver that crowns over at the outer edges and form a beam that reaches the end of the the container and  recirculates.  This then advances through the fluid until the structure reaches the ends of the container and equilibrates in a fashion akin to fig.\!~\ref{flux}.  For explanations that seek to explain this in terms of a momentum intrinsic to sound this is problematic.  The motion should then be growing at all distances simultaneously rather than working its way out.  Such an observation is consistent however with the introduction of vorticity at the edge of the driver through a ``blowing-sucking'' asymmetry as is known from pushing out and pulling in air from an orifice.  

Acoustic waves can transfer energy to walls of the container and other solid surfaces similarly to the case of ocean waves above.  This is particularly pronounced at surface irregularities which are smaller than the amplitude of motion and asymmetrically oriented to the flow.  Vorticity is created at the edge of the driver then diffuses through the fluid to the walls.  Since this requires a steady introduction of momentum from the driver there is an asymmetry in the pushing and pulling of fluid from the injection of lateral flow at the driver face that requires acceleration.  

The power delivered by the driver is split between two terms.  The power to drive the damping flow and the power to drive the waves.  Since the flows are always much slower than the sound waves, these can be separated $P_{flow}+P_{waves}$.  Flow can be easily damped with baffles and other obstructions.  In this limit, the force on the driver must average to zero.  The measure of any intrinsic momentum of sound will then be in the presence of a pressure gradient across the length of the container.

In fig.\!~\ref{flux} there is a driver sending purely rightwards propagating waves through the fluid until it meets an ``anti-reflection'' layer through which it continues to propagate to the right.  The driver creates vorticity at the corners and generates a flow with momentum that is returned at the walls of the container.  There is a pressure gradient across the container due to the reflection of the fluid at the back wall.  This is not the result of some intrinsic momentum of sound but of the fact that some momentum of the return flow is lost at the walls.  There seems to be a kind of ``stress'' on the cavity due to reflection at the walls but deeper consideration shows that this is all the result of pressure distributions at the walls that are strongly cavity shape dependent and have variations at the corners of the cavity that will not be captured by any stress terms.  

This cavity shape dependence ruins the beautiful and hopeful derivations of wave stress in derivations based on a purely local use of variables neglecting long range pressure variations or highly symmetric momentum flux analysis \cite{Lighthill78}.  The acceleration of fluid at the driver gives an asymmetric force on the driver that is balanced by backwards drag forces on the walls.  We can make a rough approximation of this for nice laminar cases but, in practice, such flows are almost always turbulent and possess a viscous boundary layer.  This means results will often be inconsistent depending on details of the cavity and driver shape and size.  Experiments bear this out and hopefully this will put an end to this chapter of overly hopeful analytic treatments of such problems.  
From a practical point of view, interesting quantities to analyze such systems are 
\begin{enumerate}
\item Driver shape and size
\item Cavity shape and size
\item Location of driver in cavity (including backside flow properties)
\item Fluid drag on walls
\item Vorticity sourcing and flux
\end{enumerate}

A prominent approach to acoustic streaming has been from Lighthill, Westervelt, and Nyborg \cite{Lighthill,Westervelt,Nyborg} whereby attenuation is the driver of of the flow.  This cannot be due to attenuation in the bulk as we now show.  We can readily see from the viscous N-S equations that eqn.\ \ref{P} is modified by a term $-\mu\nabla^{2}<(\nabla\cdot v)>$.   This gives a time average of zero for a wave driven and damped away from its source.  The resulting pressure charge averages to a decaying source with a net reduced pressure at the driver.  For an isolated driver in an infinite fluid this gives no net force on the fluid at all.  If the driver is close to a container boundary, as is typical, there is a pressure imbalance that is removed by a small net density redistribution.  These solutions are qualitatively similar of the usual solutions without damping.  Of course, viscosity and a boundary layer exists at any such driver and some streaming flow will likely result however the shape of the driver can be modified independently to the direction of vibration.  This allows acoustic and streaming motion to be directionally uncorrelated and seems like a promising test for such hypotheses.

\section{Conclusions}
Acoustics is one of the branches where hydrodynamics stress and momentum have played an enduring role yet derivations using it have been poorly validated.  Here we have shown that, in liquids, these notions are intrinsically flawed and that realistic analysis is going to be ugly for almost any system we care about.  Some parts of theoretical physics have been going through a period of heady indulgence for a number of decades.  The current insistence on experimental validation is now preeminent not just for obvious reasons but out of a frustration with a palate of derivations that are so lean and formulaic that their meaningfulness is obscured even as their mathematical correctness is verified.  This agreement should never be enough regardless.  The logical connection of the derivation to its foundations and internal consistency is also essential.  It is the author's opinion that this facet of theory has waned under publication pressures and allowed a number of such formal misapplications to persist.  

This paper is meant as a polite challenge to some long standing notions that have been promulgated by some distinguished members of the Pantheon of physics.  After a long period of experimentation on waves and ever increasing analytically powerful tools, it seems that better agreement should have been found.  Physics problems fall into the category of those solvable by clean analysis or ``clever tricks'' and those that are intrinsically ``hard.''  The first category makes good homework problems and leads most of us to believe that this is where true understanding lies.  So firm is this belief and our respect for the old greats that mathematically pretty but confused derivations can persist for a long time.  We sometimes forget that they were doing the best they could with the mathematical and computational tools they had and that some of their clever constructions might be simply wrong.  Hopefully, the arguments here will convince people to abandon these and seek more direct solutions that consider boundary effects properly.   

The nonlinear correction to the pressure field is easy for highly symmetrical systems but these are often cases where we cannot make easy measurements.  One possible exception is in the highly symmetric implosions of sonoluminescence.  The bulk viscosity is nearly zero for liquids but the nonlinear pressure charge $\rho_{P}$ creates an additional pressure drop about the collapsing bubble.  For the very high velocities obtained here, this may be an important contribution.  Acoustic levitation and streaming have practical importance.  A knowledge of the relevant macroscopic features and well founded salient examples are crucial in driving a field forwards.  Hanging on to old imperfect notions of hydrodynamic stress will only foment confusion and blur validity of derivations in a subject where the topic is classical enough that it should really be on a firmer footing.  

\section{Appendix}
\begin{figure}[!h]
   \centering
   \includegraphics[width=3in,trim=0mm 160mm 0mm 160mm,clip]{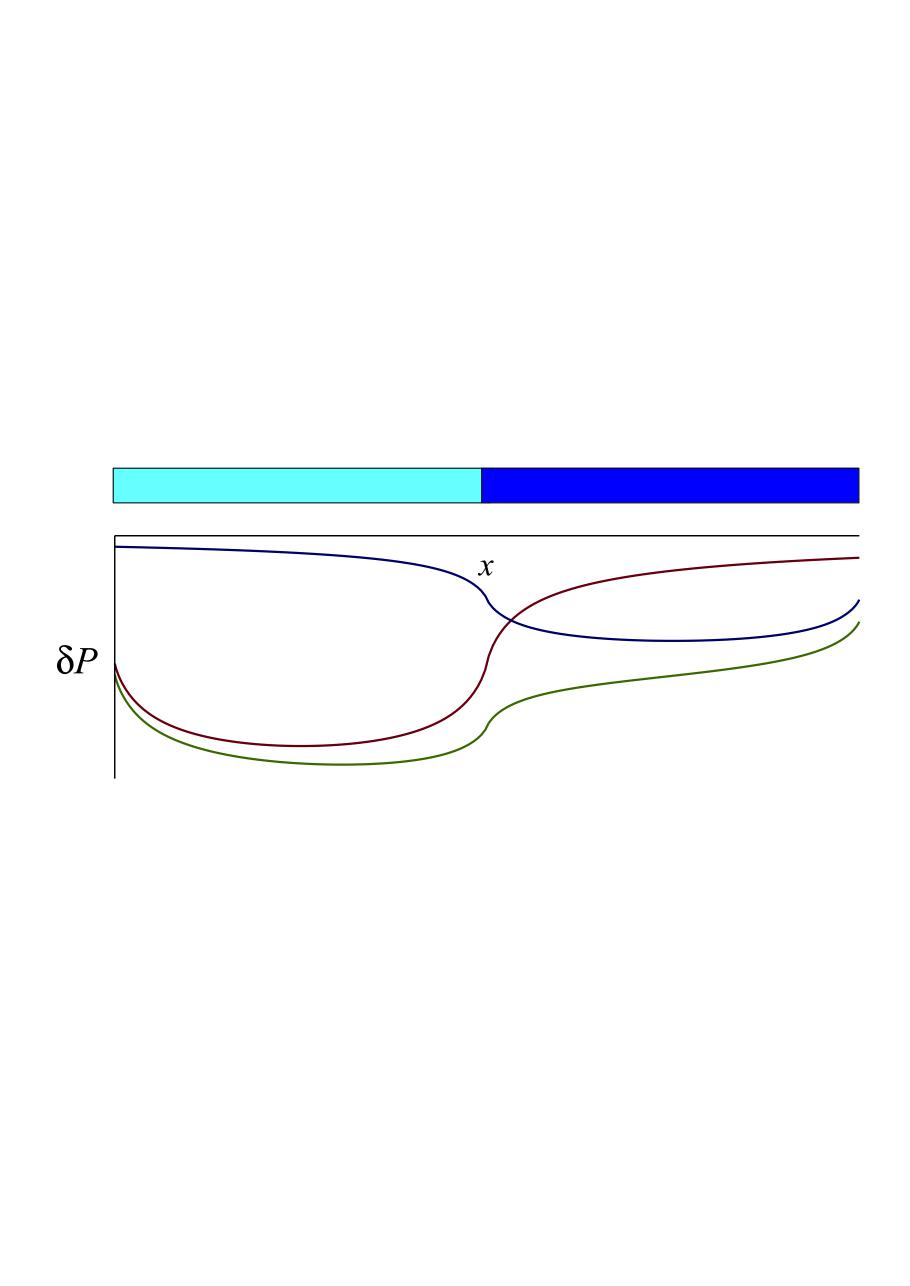} 
   \caption{Initial pressure variation in the strip: first fluid (red line), second fluid (blue), net (green). }
   \label{pressure}
\end{figure}

\begin{appendix}
Here were will consider the case of a 2D strip of two adjacent liquids of lengths $l$ in a thin container of width $d\ll l$.  The volume is fixed and the liquids are at a pressure $P_{0}$ before end actuators generate a sound wave of mean ``pressure charge'' $c_{1}$ in the first and $c_{2}<c_{1}$ in second.  The resulting initial change in the pressure in each segment can be computed by Green's functions
\begin{align}
\delta P=\int_{0}^{l}\left(-\frac{1}{4\pi}\right)\frac{1}{|x-x'|}c_{1}d'x+\int_{l}^{2l}\left(-\frac{1}{4\pi}\right)\frac{1}{|x-x'|}c_{2}d'x
\end{align}
and the result upon adding in shown in fig.\!~\ref{pressure}.  
When these long range pressure fluctuations settle down we assume the new density changes are small so that the distribution of sound waves remain relatively unaffected.  Mass conservation yields a new mass density 
\begin{align}
\rho(x)=\rho'-\beta\rho_{0}\delta P(x)
\end{align}
where $\int \rho=\rho_{0,L}~\text{Vol}_{L}+\rho_{0,R}~\text{Vol}_{R}=M_{\text{net}}$ so the resulting net $P(x)$= const.   The interface will get a net displacement leftwards proportional to the volume change from the bulk equilibration of the mass under the acoustic field.  
\end{appendix}

\end{document}